# Dynamics of Urban Heat Island in Lafia, Nasarawa State of Nigeria: A Remote Sensing Analysis of Land Surface Temperature, Urban Development and Vegetation Change

*Oladiran Johnson Abimbola, Taiwo Adewumi and Musa Abubakar

Department of Physics, Faculty of Science, Federal University of Lafia, Nigeria

*Corresponding Author: abimbola.oladiran@science.fulafia.edu.ng

## Abstract

As the global climate changes, urban heat island (UHI) is a critical factor in ever expanding urban landscape, studying and mitigating the UHI is important for remediating climate change and providing for the human and ecosystem health within the urban area. This study has aimed to study the UHI in Lafia, a tropical city in Nigeria and its other impacted factors such as the land surface temperature (LST) and normalized difference vegetation index (NDVI), with the aim of mitigating the UHI effect. Landsat 4, 5, 7 and 8 together with Sentinel data has been used for this study, through the public archive of the Google Earth Engine data catalog, used also is the ERA5 data from the same data catalog. The result showed that the expanding city of Lafia is experiencing significant UHI with increase in temperatures in the city and adjoining areas, it was found that the vegetation cover in Lafia city is rapidly disappearing as a result of urbanization leading to more UHI and greater discomfort to the inhabitant of the city. Several remediation steps were suggested to mitigate the UHI effect in Lafia.

**Introduction**

The Urban Heat Island (UHI) phenomenon describes the observed temperature increase in Urban areas as compared to their rural surrounds. This temperature disparity is caused by anthropogenic activity, changes to natural landscapes, and the thermal characteristics of urban materials. UHIs are important in urban climatology because they greatly affect environmental quality, energy consumption, and human health in densely populated places (Oke *et al.*, 2017).

Natural surfaces like rivers, water bodies, and vegetation are typically destroyed during the urbanization process to make way for roads and buildings made of concrete, metal, asphalt, and other highly thermally active materials. These materials typically have low albedo and high heat capacity, which results in high absorptivity and retention of solar radiation during the day and slow emission during the night (Mohammad and Goswami, 2021). This causes the nighttime temperatures in urban areas to be abnormally warmer than those in nearby rural areas with more vegetation.

The vegetation index, particularly the Normalized Difference Vegetation Index (NDVI), which measures plant density, is directly proportional to the extent of the UHI influence. According to studies (Bijesh *et al.*, 2019; Feyi & Ibeabuchi, 2023; Zheng *et al*., 2023; Aktaviani *et al.*, 2024), metropolitan areas with low NDVI values experience more acute UHI due to less shade and vegetation-induced cooling. Increasing evaporative cooling and expanding the area of water bodies are two ways to mitigate the effects of UHI (Saaroni et al., 2018; Susca & Pomponi, 2020; Dionysia, 2022; Guglielmina *et al*, 2024). The amount of atmospheric moisture brought on by evapotranspiration is often measured using the Normalized Difference Water Index (NDWI). Differential heat stress and localized microclimates will arise from the uneven spatial distribution of vegetation and water bodies inside the metropolitan area.

Land surface temperature (LST) and changes in atmospheric conditions are critical factors that drive UHI effect seasonally across tropical regions including West Africa, meanwhile, moisture in the atmosphere with rainfall increase during the raining seasons help to lower the atmospheric temperature and thereby reduced the effect of UHI. During the dry season, which includes the harmattan periods, as a result of considerable reduction in atmospheric moisture, UHI is enhanced with the urban area having higher land surface temperature than the rural area. Besides, the urban-rural vegetation cover, oscillating between wet and dry seasons is also a significant factor in influencing the UHI effect.

Urban heat island effect has significant effects on our environment: increase in urban temperature will lead to more energy demands require for operating cooling devices, consequently leading to more greenhouse gas productions and financial burden on individuals; severe UHI can lead to heatwaves in the urban area with the attendant grave consequence for vulnerable people such as the infants and the elderlies, including very sick people. In addition to affecting human health, the UHI effect has a negative impact on urban biodiversity, as species that are less heat-tolerant struggle to adapt to higher temperatures in urban contexts.

The UHI phenomena highlights the interaction between urbanization and microclimate changes. Water bodies and vegetation cover greatly influenced environmental thermal settings and this is demonstrated by the regional and seasonal variations in UHI intensity. It is pertinent that scientific research knowledge together with proper and sustainable urban planning be used to mitigate the problems posed by UHI and our changing climate.

## Methodology

**Study Area**

Lafia is a capital city of Nasarawa State, within the north-central region of Nigeria. Lafia, as a region of interest (roi), for this study as shown in Figure 1, the region of interest is situated within latitudes 8.5500 $^oE$ and 8.4658 $^oE$ as well as longitude 8.6169 $^oN$ and 8.4661 $^oN$, with a total land area of 155 $km^2$. The city became the capital city of Nasarawa State in 1996 upon the creation of the Nasarawa State, prior to the time of creation of Nasarawa State, Lafia was a rural area with a human population of less than twelve thousand. Figure 2 depicts the human population of the roi for the years 2000, 2005, 2010, 2015, and 2020 based on data from the SEDAC (2024) Population Estimation Service. It is estimated that the population of the roi has grown by 76% since 2000.

Lafia is located in the tropical savannah and has a Köppen climatic classification of *Aw* (Abimbola *et al*., 2023). According to Abimbola *et al.* (2023), the city experiences approximately 251 rainy days, 1,645 *mm* of annual rainfall, and an average relative humidity of 75%. The average temperature is 26.7 $^oC$. This demonstrates that Lafia and the surrounding area are hot, humid, and have distinct wet and dry seasons.

According to the data from the NASA Socioeconomic Data and Applications (SEDAC, 2024) and has shown on Figure 2, the human population of the region of interest has been steadily increasing overtime: it increases from about 15,905 in the year 2000 to about 28,034 in the year 2020!

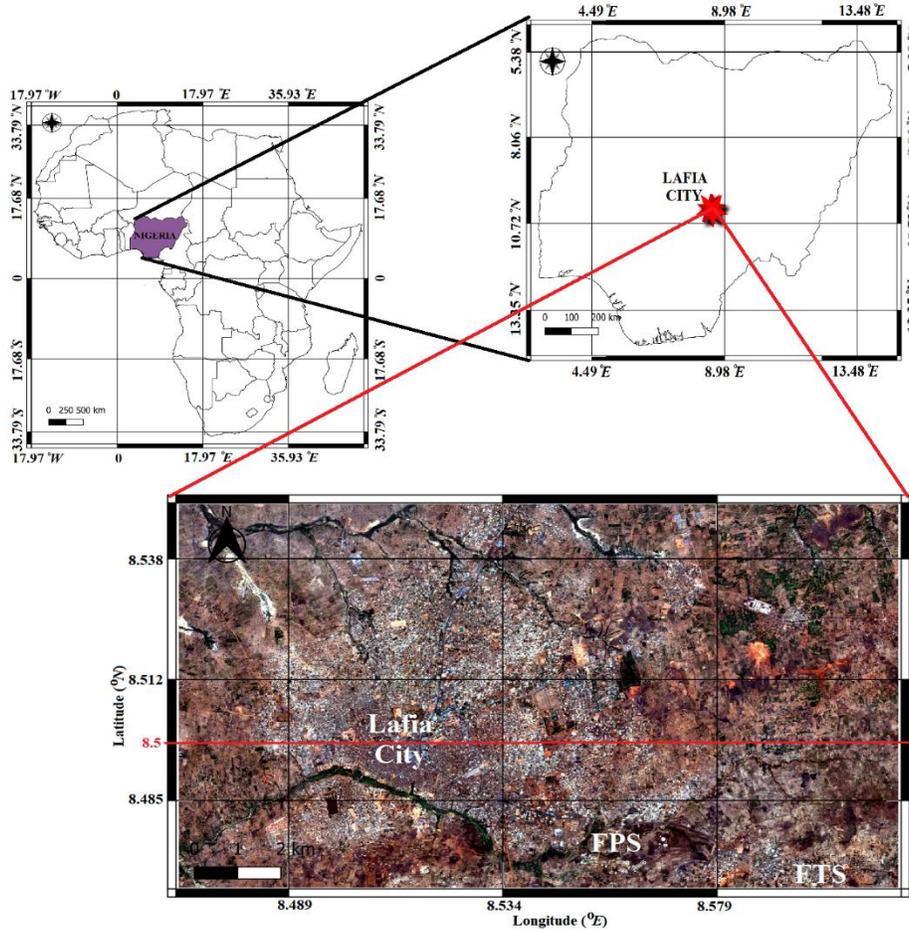

**Figure 1:** Study's Region of Interest (roi) (FPS => FULafia Permanent Site; FTS => FULafia Take-off Site)

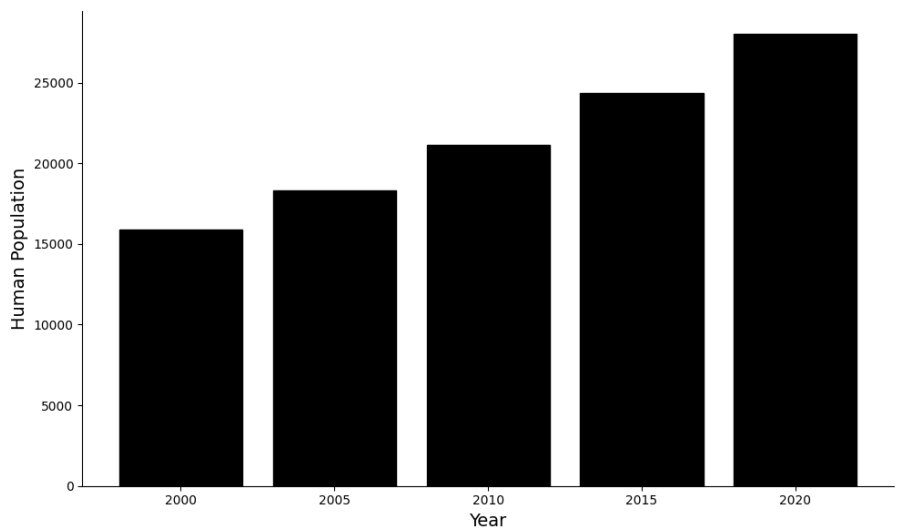

**Figure 2**: The human population growth within the region of interest (SEDAC, 2024)

**Remotely Sensed Satellite Data**

Using the data repository of the Earth Engine Data Catalog of Google at https://developers.google.com/earth-engine/datasets. Four Landsat imageries, that is, Landsat 4, 5, 7 and 8, including Sentinel-2 were used, the summary of the details of these set of imageries are shown in Table 1. The Spatial resolutions of Landsat images is 30 *m*, while that of Sentinel-2 is 10 *m*. In this study, Landsat 4 and 5 are used for 1984 to 1999 analysis, Landsat 7 was used for 2000 to 2012, while Landsat 8 was used for 2013 to 2024. Sentinel-2, Dynamic World was used for the land cover/land use (LCLU) analysis which covers the years 2015 to 2024: more information on the Dynamic World data can be found in the work of Brown *et al*. (2022).

**Table 1:** Characteristics of the satellite images used in the study

| Satellite Platform | Sensor | Path | Row | Cloud Cover | Coverage Year |
|---|---|---|---|---|---|
| Landsat 4 | TM | 188 | 54 | 10 | 1982 – 1993 |
| Landsat 5 | TM | 188 | 54 | 10 | 1984 – 2012 |
| Landsat 7 | ETM+ | 188 | 54 | 10 | 1999 – 2021 |
| Landsat 8 | OLI | 188 | 54 | 10 | 2013 – till date |
| Sentinel-2 | L1C | - | - | - | 2015 – till date |

**TM**: Thematic Mapper, **ETM+**: Enhanced Thematic Mapper Plus, **OLI**: Operational Land Imager
**L1C => Level-1C**

ERA5-Land Hourly dataset from the Earth Engine Catalog, provided by the Copernicus Climate Data Store, is used to analyze the yearly trend of land surface temperature across the study region: ERA5-Land data is available from 1950 to present on hourly temporal resolution.

Normalized Difference Vegetation Index (NDVI) which measures vegetation cover and health is calculated from Landsat image bands as (Sobrino *et al*., 2004):

$$NDVI = \frac{NIR - RED}{NIR + RED} \qquad (1)$$

where, NIR = Near Infra-Red Band (for Landsat 4, 5 and 7, this is band SR_B4, while for Landsat 8 it is SR_B5); RED = Red band (for Landsat 4, 5 and 7 it is SR_B3, while for Landsat 8 it is SR_B4).

The brightness temperature (BT) is calculated thus:

$$BT = T\_B \times 0.00341802 + 149.0 \tag{2}$$

where, T_B is the Therma Band (for Landsat 4, 5 and 7 this is ST_B6, while for Landsat 8 it is ST_B10).

The land surface temperature (LST) in *Kelvin*, is then calculated using Equation 3:

$$LST = \frac{BT}{1+\left(\frac{\lambda \cdot BT}{\rho}\right)\cdot \ln(\epsilon)} \tag{3}$$

where, $\lambda$ = is the emitted radiance wavelength (for Landsat 4, 5 and 7 it is 11.5 $\mu m$, while for Landsat 8 it is 10.9 $\mu m$); $\rho = h \cdot c/\sigma$ ($h$ = Planck's constant, $c$ = speed of light = $2.998 \times 10^8$ m/s, and $\sigma$ = Boltzmann constant = $1.38 \times 10^{-23}$ J/K); $\epsilon$ is the land surface emissivity which is calculated as:

$$\epsilon = 0.004 \times fvc + 0.986 \tag{4}$$

$fvc$ is the fraction of land covered by vegetation and it is estimated from NDVI as:

$$fvc = \left(\frac{NDVI - NDVI_{min}}{NDVI_{max} - NDVI_{min}}\right)^2 \tag{5}$$

$NDVI_{max}$ and $NDVI_{min}$ are maximum and minimum values of NDVI.

The urban heat island (UHI) is calculated as:

$$UHI = \frac{BT - \overline{BT}}{\overline{BT}} \tag{6}$$

with $\overline{BT}$ is the average of the brightness temperature.

## Results

### Land Surface Temperature (LST)

Figure 3 shows the distribution of land surface temperature (LST) within Lafia and its environment for the 40-year period interval of 1984 to 2024; it clearly shows the temporal evolution of thermal conditions within Lafia. The mean LST, as shown in Figure 3, ranges between 34.2 °C, represented by a blue colour, to 47.5 °C, represented by a red colour. Indicated on Figure 3 are key locations such as Lafia City, Federal University of Lafia (FULafia) permanent site (PS), and FULafia take-off site (TS) for spatial context.

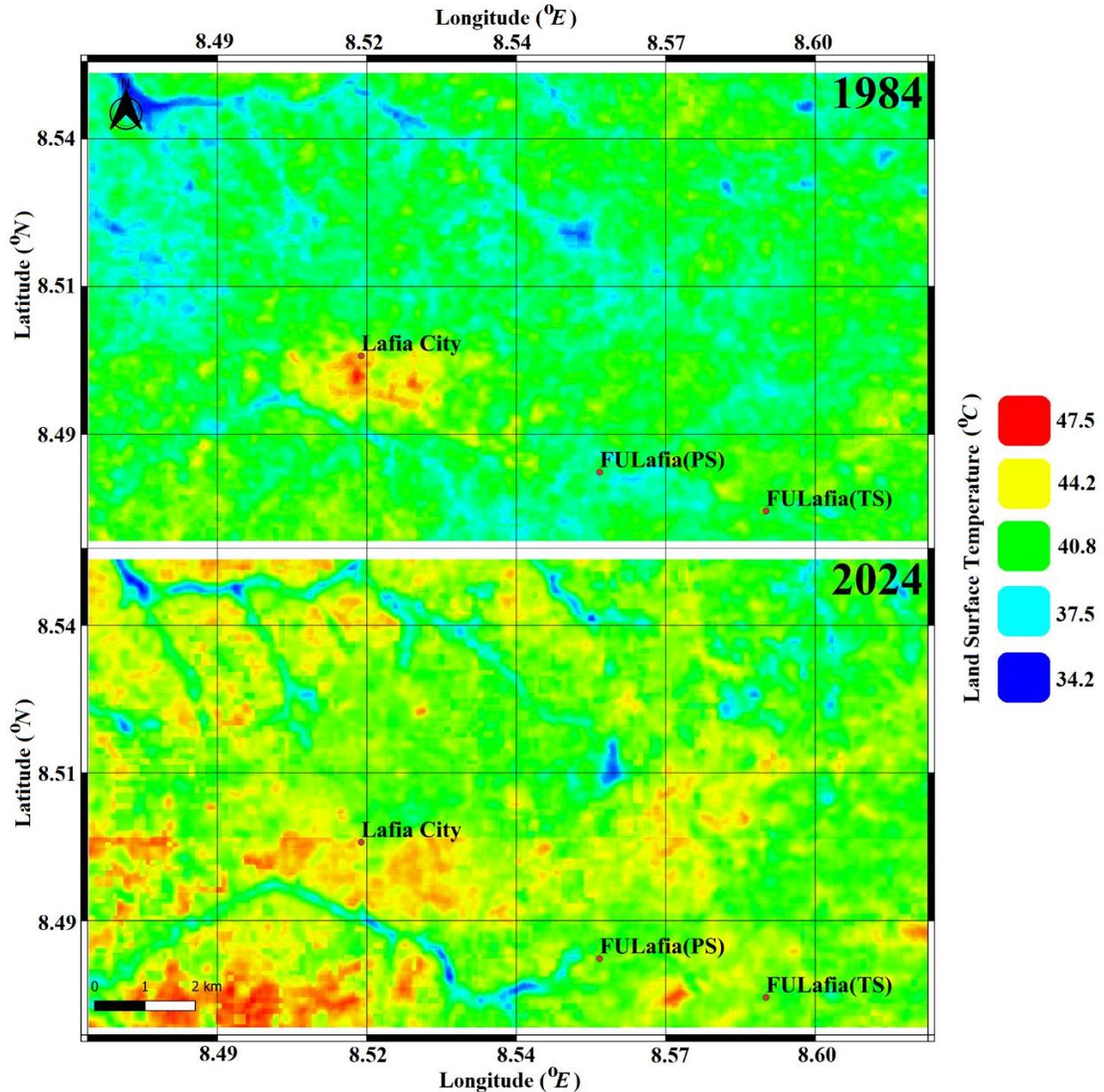

**Figure 3:** Land surface temperature (LST) in Lafia and its surroundings for the years 1984 and 2024.

The temperature profile along the line of latitude 8.5 °*North* are shown on Figure 4. As shown on Figure 1, latitude 8.5 °*North* passes through the Lafia city into the rural environment around the Lafia city: temperature profiles for the years 1984 and 2024 are shown.

Annual mean temperature and the temperature anomaly taking the 20$^{th}$ century regional average (1991 to 2000) as a baseline, from Copernicus ERA5 data for the year 1980 to 2024, are shown on

Figure 5: the 1991 - 2000 regional average used in this work is 27.56 °C. The linear fit on the annual temperature distribution gives the correlation equation of Equation 7:

$$T = 0.023Y + 27.32 \qquad (7)$$

where, $T$ is the annual mean land surface temperature (°C) and $Y$ is the year in question (1980 to 2024).

The red bars of the anomaly plot show the years with land surface temperature above the regional 20th century average while the green bars show the years with temperature below the regional 20th century average.

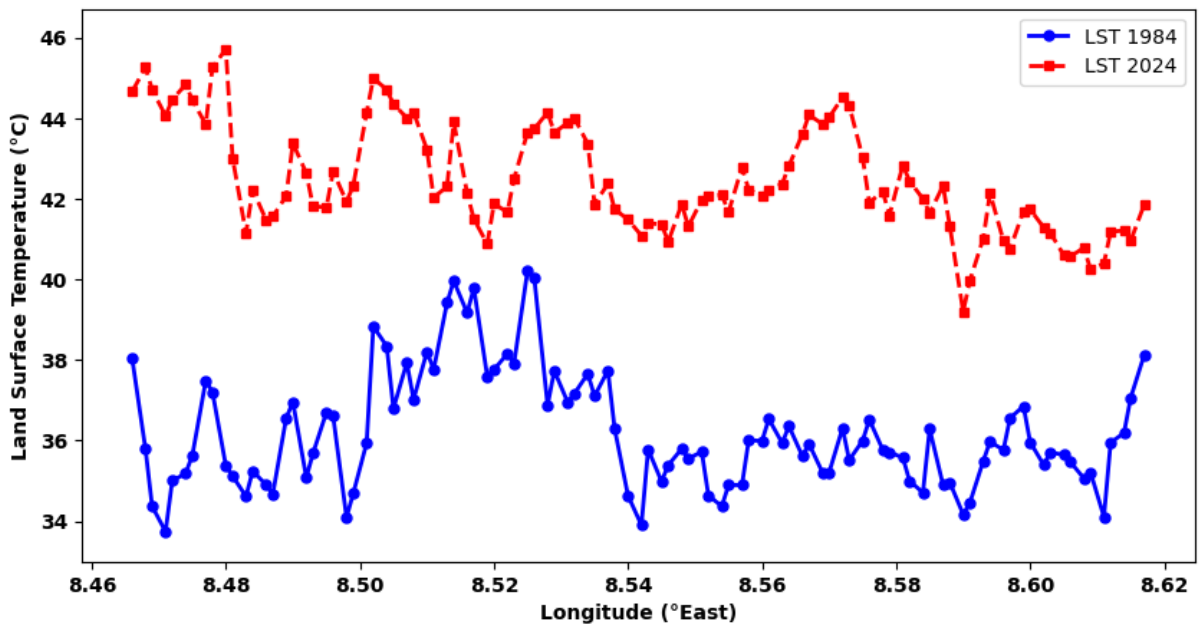

**Figure 4:** Land Surface Temperature profile along latitude 8.5 °North for the years 1984 and 2024

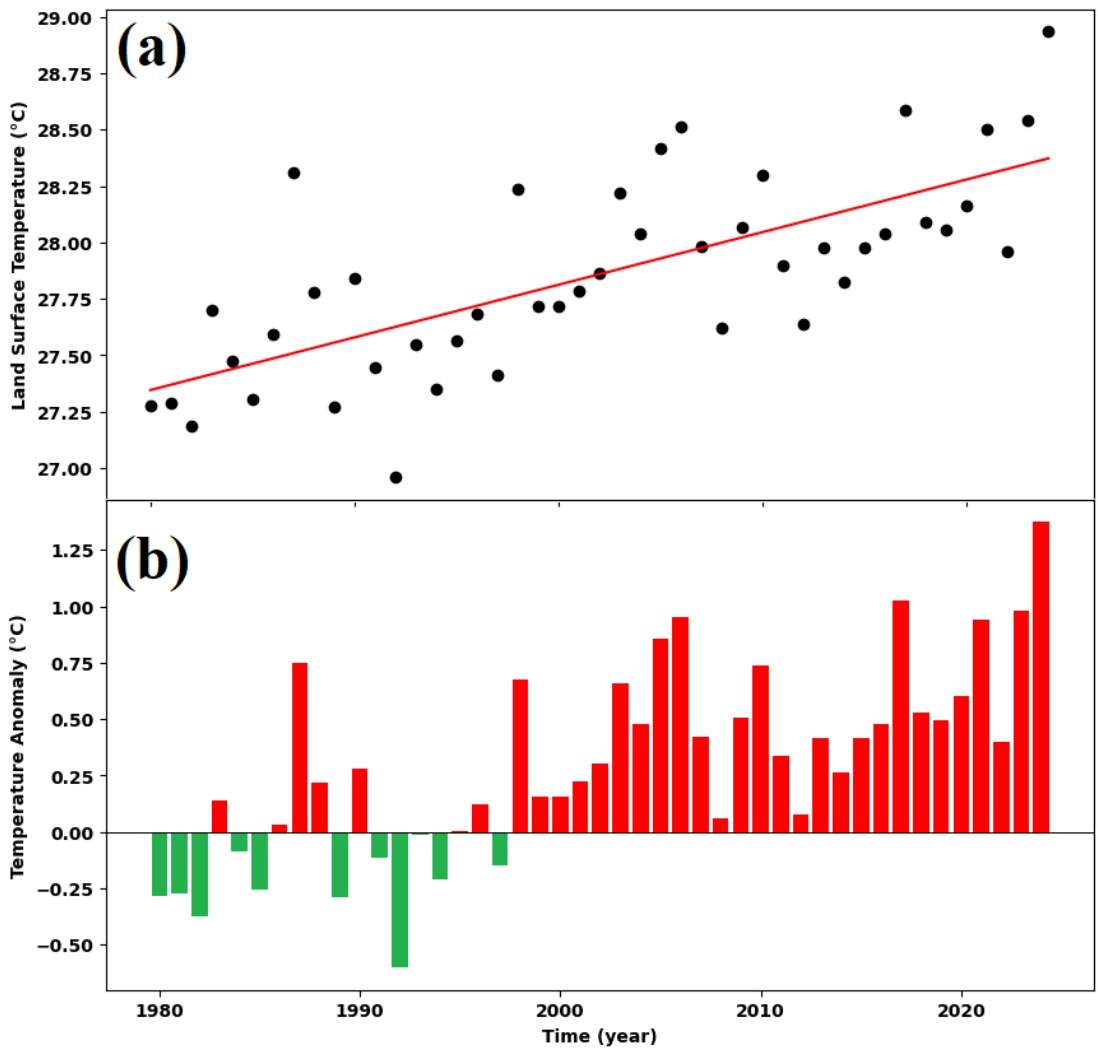

**Figure 5:** Region of interest temporal temperature variation: (a) Temperature distribution from 1980 to 2024, (b) Temperature anomaly for the year 1980 to 2024

**Normalized Difference Vegetative Index (NDVI)**

Normalized difference vegetation index (NDVI) which is normally used to indicate vegetation health and greenness is shown in Figure 6, for the region of interest, for the year 1984 and 2024. Latitude 8.5 $^o$*North* spatial profile of the NDVI within the region of interest is also shown on Figure 7, for the years 1984 and 2024. The value of NDVI within this region ranges between 0.1 and 0.6:

it should be noted that the Federal University sites were not in existence in 1984!

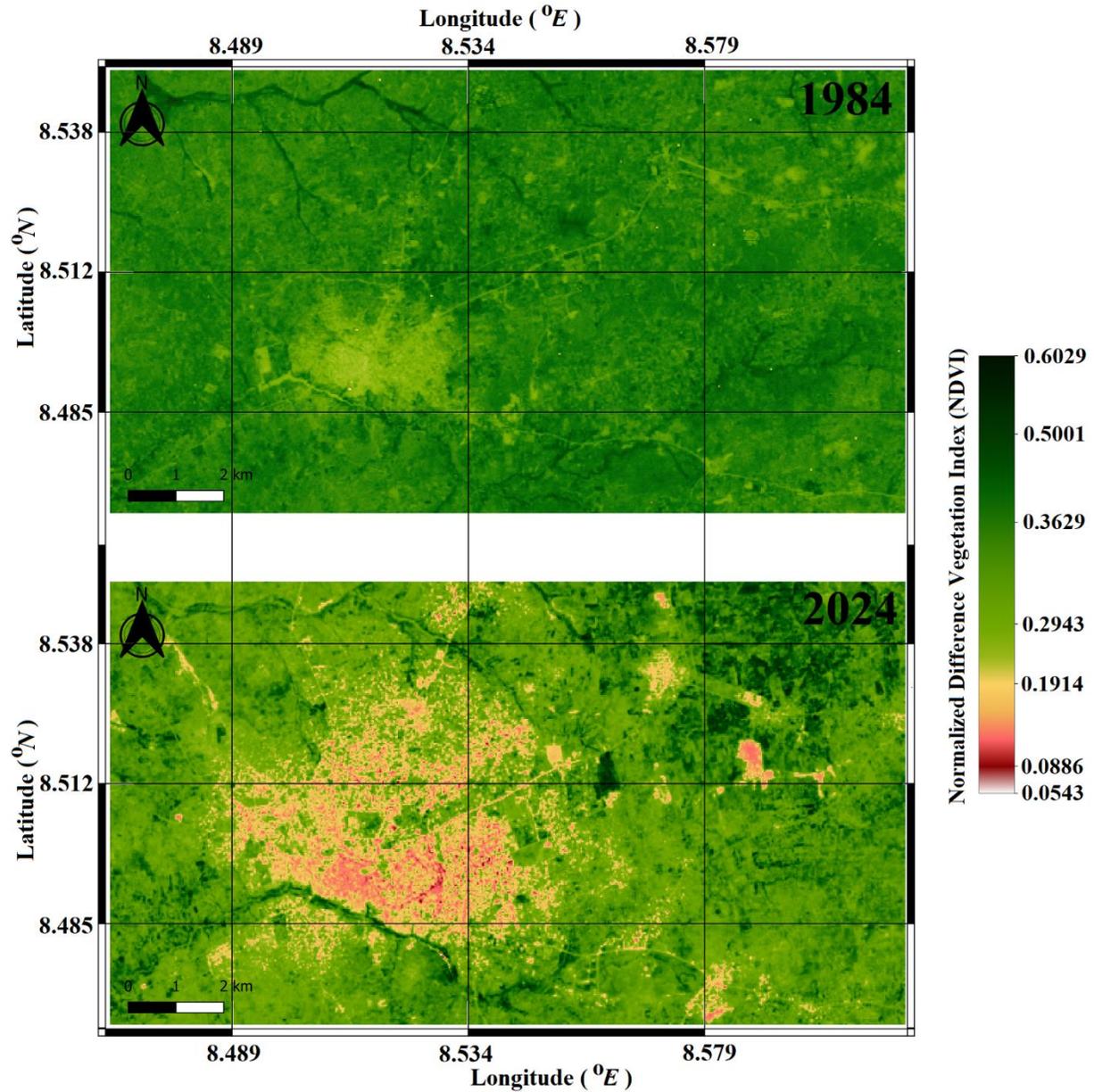

**Figure 6:** Normalized Difference Vegetation Index (NDVI) for the year 1984 and year 2024, for the region of interest

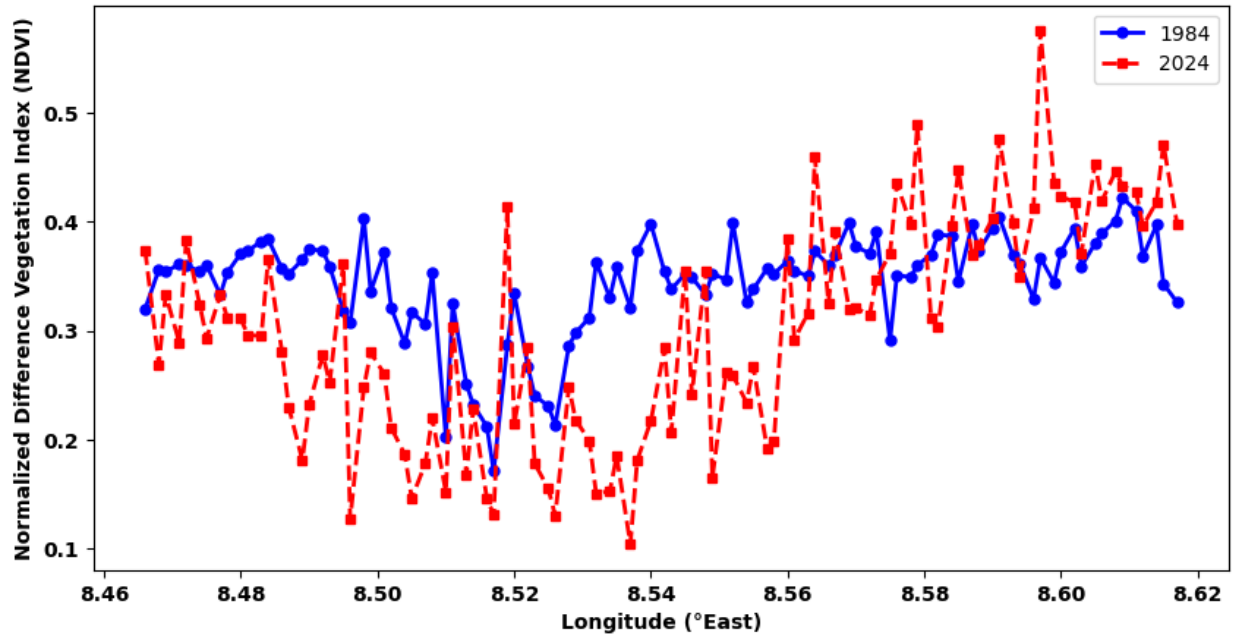

**Figure 7:** Profile of the Normalized Difference Vegetation Index along the latitude 8.5 $^oNorth$, for the years 1984 and 2024

**Land Use Land Cover (LULC)**

Using Sentinel data, the land use/land cover classifications for the region of interest is presented on Figure 8, for the year 2015 and 2024. Figure 8 shows the classes of land use such as water body, trees, grass cover, flooded vegetation area, crops or farmland, shrubs and scrub, built up area, and bare ground. It should be noted that some of these classes may actually overlap: for instance, satellite may see rice field as grass covered land, while a rocky surface may be classed as a built-up area!

The land cover changes in terms of squared kilometer and percentage of the total area under consideration is shown on Table 2 which is prepared according to the classes of LULC for the years 2015, 2018, 2021 and 2024.

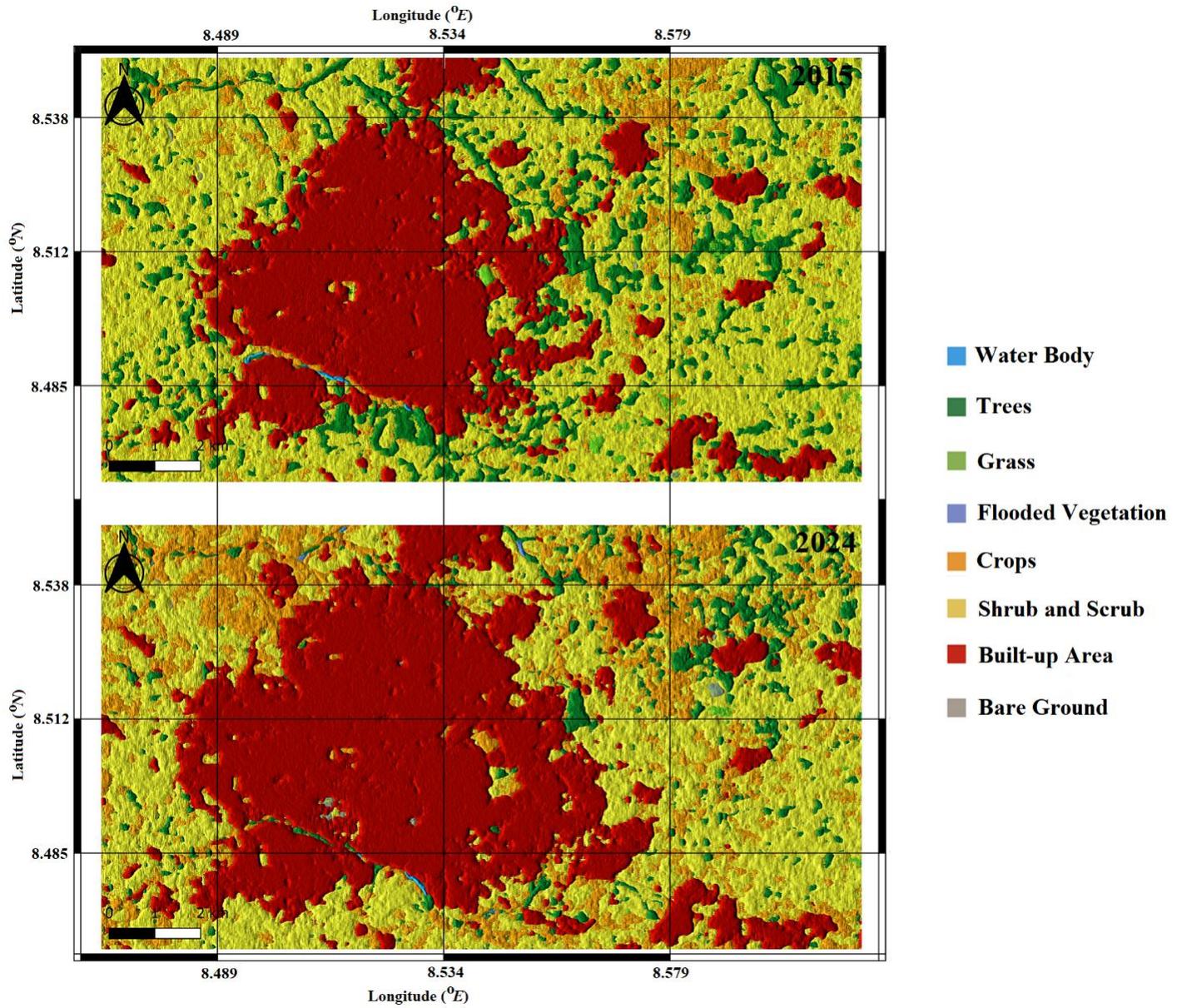

**Figure 8:** Land use land cover (LULC) changes between the year 2015 and 2024

Table 2: LULC variations for the year 2015, 2018 2021 and 2024 across the region of interest.

| | | 2015 | 2018 | 2021 | 2024 |
|---|---|---|---|---|---|
| Built-up | (km²) | 46.53 | 44.25 | 51.32 | 61.87 |
| | (%) | 30.10 | 28.62 | 33.20 | 40.02 |
| Water | (km²) | 0.12 | 0.08 | 0.02 | 0.07 |
| | (%) | 0.08 | 0.05 | 0.01 | 0.04 |
| Trees | (km²) | 21.82 | 3.36 | 12.36 | 9.23 |
| | (%) | 14.11 | 2.17 | 8.00 | 5.97 |
| Grass | (km²) | 0.82 | 0.70 | 0.39 | 0.17 |
| | (%) | 0.53 | 0.45 | 0.26 | 0.11 |
| Crops | (km²) | 11.44 | 45.29 | 13.28 | 26.74 |
| | (%) | 7.40 | 29.29 | 8.59 | 17.30 |
| Shrub-Scrub | (km²) | 73.72 | 60.54 | 76.85 | 56.25 |
| | (%) | 47.68 | 39.16 | 49.71 | 36.38 |
| Flooded Vegetation | (km²) | -- | 0.01 | 0.08 | 0.08 |
| | (%) | -- | 0.01 | 0.05 | 0.05 |
| Bare Ground | (km²) | 0.16 | 0.38 | 0.31 | 0.20 |
| | (%) | 0.11 | 0.25 | 0.20 | 0.13 |

**Urban Heat Island (UHI)**

The Urban Heat Island spatial distribution with the region of interest is shown on Figure 9, for the years 1984 and 2024. The values of the UHI are indicated with the color code showing red color as the highest UHI (0.026) while blue indicates lowest UHI (-0.021).

The spatial distribution of the UHI along the latitude line of 8.5 °*North* is shown on Figure 10, for the same years of 1984 and 2024.

The class of the UHI was formed following the work of Naim & Kafy (2021) into Weak (green), Middle (blue), Strong (orange), Stronger (brown) and Strongest (red), with some very small area that cannot be classed! Tables 3 and 4, shows the yearly variation of these classes across the region of interest for the years 1984, 1991, 2003, 2014 and 2024, squared kilometer and percentage.

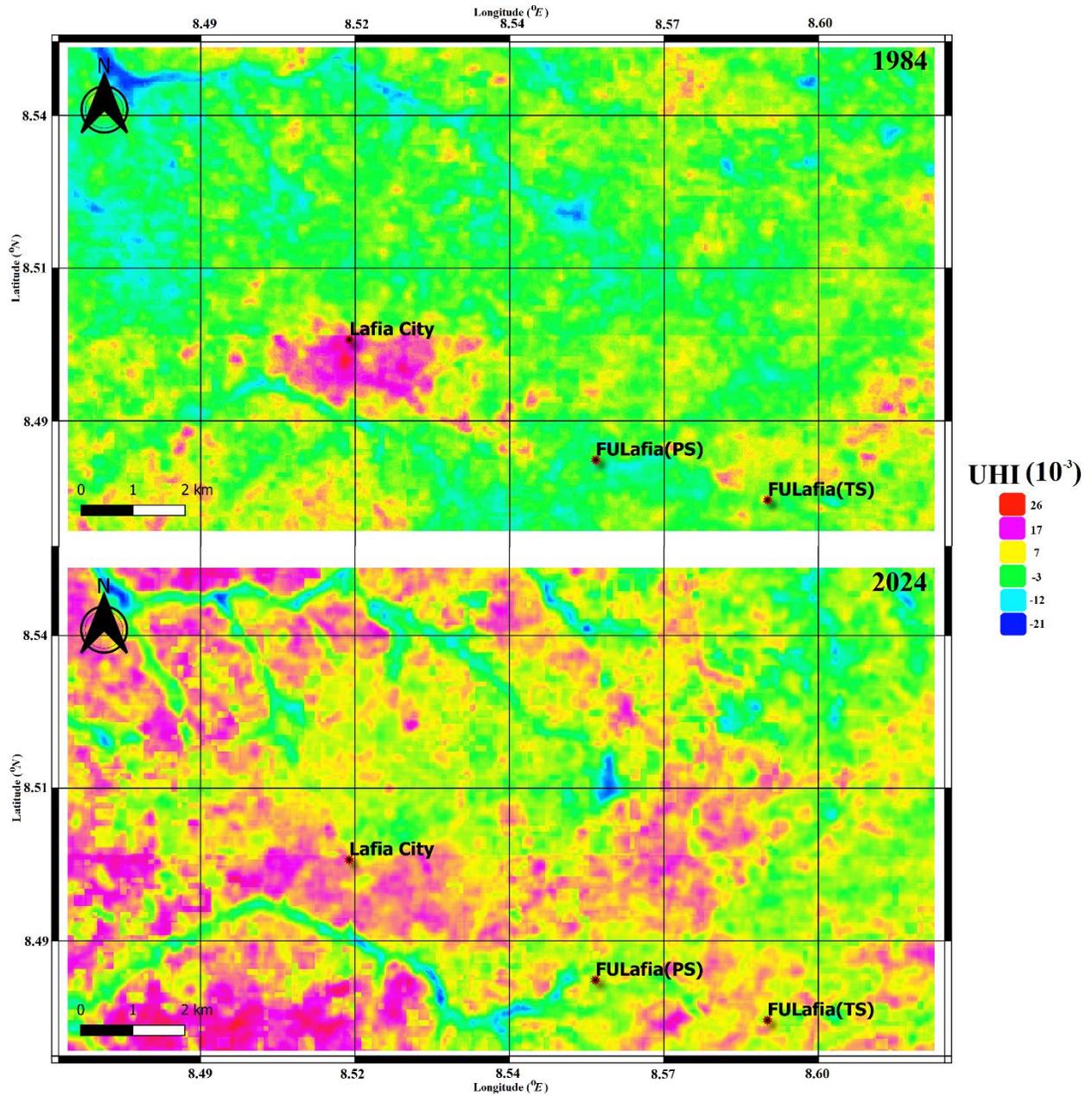

**Figure 9:** Spatial distribution of Urban Heat Island within the region of interest for the years 1984 and 2024

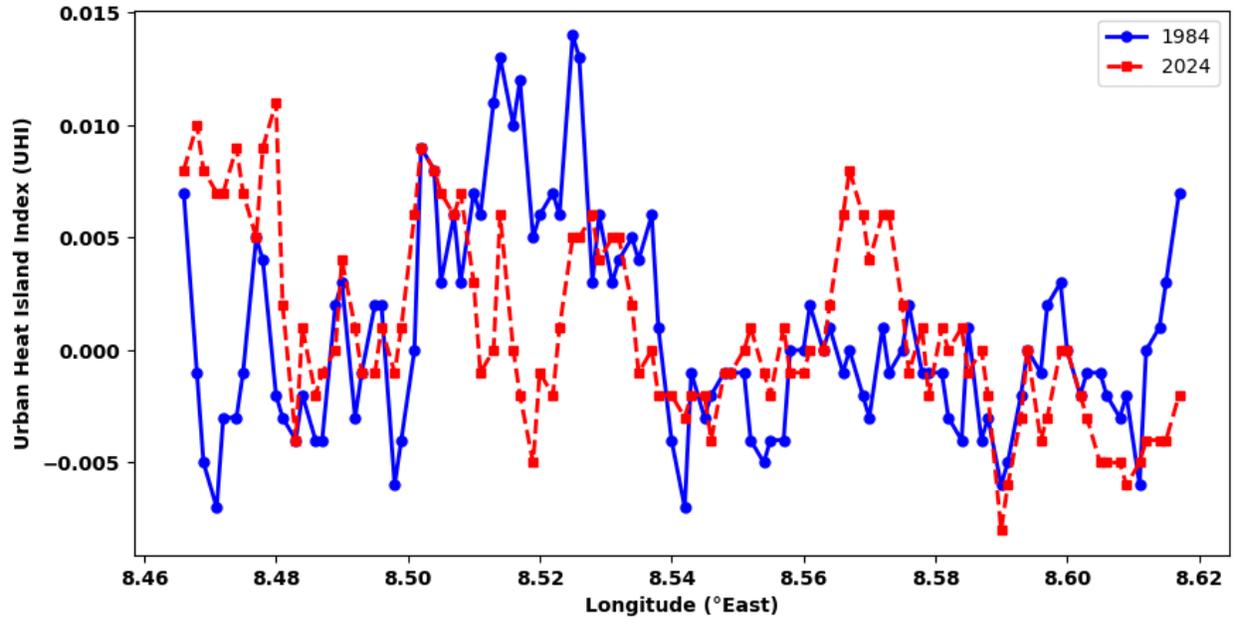

**Figure 10:** Spatial variation of Urban Heat Island Index along the line of latitude 8.5 *ºNorth*

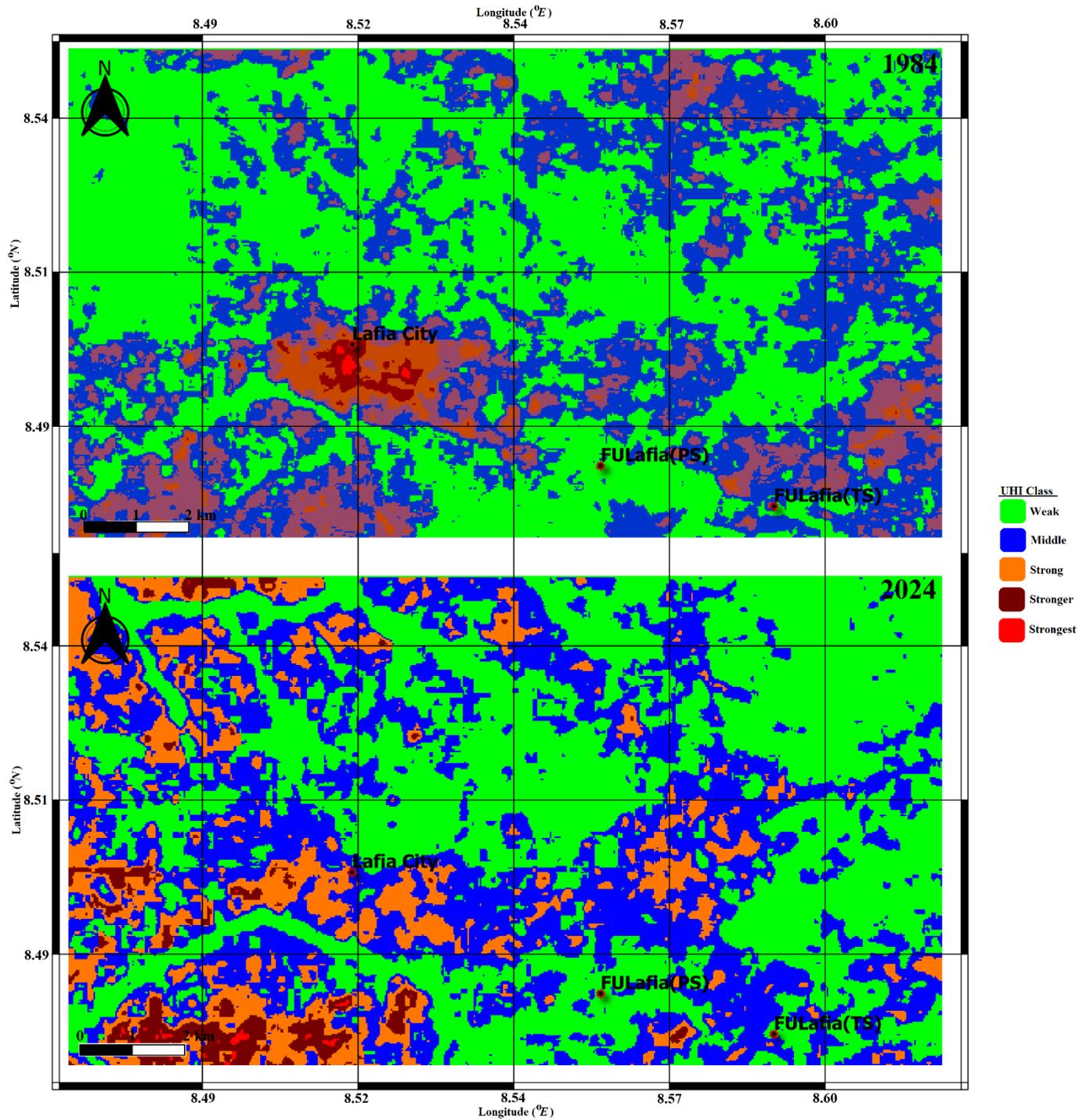

**Figure 11:** Urban Heat Island, according to class distribution within the region of interest for the years 1984 and 2024

Table 3: Total Area of Land Covered by the Classes of the Urban Heat Island of the total land area under consideration

| Class of UHI | 1984 (km²) | 1991 (km²) | 2003 (km²) | 2014 (km²) | 2024 (km²) |
|---|---|---|---|---|---|
| Weak | 79.57 | 78.52 | 78.45 | 76.97 | 74.37 |
| Middle | 54.29 | 56.49 | 38.51 | 48.92 | 56.96 |
| Strong | 16.76 | 17.12 | 26.19 | 23.43 | 19.54 |
| Stronger | 3.39 | 2.76 | 10.14 | 4.81 | 4.20 |
| Strongest | 1.04 | 0.28 | 1.76 | 0.88 | 0.13 |
| Undefined | 0.13 | 0.01 | 0.14 | 0.18 | -- |
| TOTAL | **155.18** | **155.18** | **155.18** | **155.18** | **155.18** |

Table 4: Percentage Land Cover by the Classes of the Urban Heat Island of the total land area under consideration

| Class of UHI | 1984 (%) | 1991 (%) | 2003 (%) | 2014 (%) | 2024 (%) |
|---|---|---|---|---|---|
| Weak | 51.27 | 50.60 | 50.55 | 49.60 | 47.92 |
| Middle | 34.99 | 36.40 | 24.82 | 31.52 | 36.70 |
| Strong | 10.80 | 11.03 | 16.88 | 15.10 | 12.59 |
| Stronger | 2.18 | 1.78 | 6.54 | 3.10 | 2.70 |
| Strongest | 0.67 | 0.18 | 1.13 | 0.57 | 0.09 |
| Undefined | 0.09 | 0.01 | 0.08 | 0.11 | -- |
| TOTAL | **100** | **100** | **100** | **100** | **100** |

**Discussion**

**Land Surface Temperature (LST)**

The land surface temperature (LST) result shows an increase in surface temperature across the study area over the study period of 1984 to 2024 (a 40-year period), the average surface temperature was found to range between 34.2 $^oC$ and 47.5 $^oC$. The urban area of Lafia city as well as places such as the permanent (PS) and take-off (TS) sites of the Federal University of Lafia, which was established in the year 2011, show elevated surface temperature, as presented on Figure

3. It can be observed that in 1984, Lafia city is the only area with temperature around 47.5 $^oC$, all other areas within the study region show lower average temperature, but 40 years after in the year 2024, the average temperature across the study region has gone up and average temperature of 47.5 $^oC$ has become a common place. The upward trend in surface temperature over the years, as observed in Lafia is agreement with the observed global upward trend in urban and developing urban areas due to changes in land use that is replacing natural vegetations with concrete and asphalt surfaces (Oke, 1982; Zhou *et al.*, 2014).

Urban-rural temperature contrast is a well researched phenomenon in the study of urban heat island (Arnfield, 2003) and this same effect can be clearly seen on Figure 4, where the spatial surface temperature profile along the line of latitude 8.5 $^oNorth$ is shown. On the profile for 1984, an elevated urban temperature can be observed between longitude 8.50 and 8.54 $^oEast$, this clearly shows the effect of UHI around Lafia in 1984. By the year 2024, the distinction between urban area and the rural area is becoming blurring as the general temperature of the area is increasing in response to anthropogenic climate change effect.

The annual mean temperature of the study area was found to have been increasing steadily, with significant increase observed to have started around the year 2000, as shown on Figure 5(b) and it can be seen that the year 2024 shows the highest temperature increase when compared with the 20th Century (1991 – 2000) regional average: the upward trend in temperature anomalies show the influence on local climate by the process of urbanization, urban area is generally known to absorb and retain thermal energy than the adjoining rural area (Oke, 1982). The linear fit on Figure 5(a), which results in Equation 7, shows that within the study period, the regional temperature has been increasing at the rate of approximately 0.023 $^oC$ per year! This increase in surface temperature due to urbanization of Lafia is in line with the observation of other researchers across the globe working on similar sprawling cities similar to Lafia (Ayanlade *et al.*, 2017; Abulibdeh *et al.*, 2019).

**Vegetation Cover**

Figure 6 show significant decline in vegetation cover between 1984 and 2024, this decline is clearly evident around the Lafia urban area and immediate environ and some rural areas that were covered in vegetation can be seen to have been deprived of vegetation in the year 2024, 40 years after: the urbanized area could be observed to have a very low NDVI value of around 0.089 while the rural area have NDVI value of around 0.60. Vegetation cover is a primary component for moderating land surface temperature through shading and evapotranspiration, hence the reduction in vegetation cover, as observed, is a key driver of urban heat island in Lafia, this is also corroborated by the work of Buyantuyev and Wu (2010).

The NDVI spatial profile along the line of latitude 8.5 $^oNorth$, as shown on Figure 7, shows clear differences between urban and rural vegetation cover: in 1984, the clear low NDVI value between longitude 8.5 $^oEast$ and 8.54 $^oEast$, which correspond to the region with elevated surface temperature, as shown in Figure 4. By the year 2024 the NDVI values around this region has even gone lower, indication greater reduction in vegetation, meanwhile some rural areas to the east are found to have somewhat increased vegetation cover. The effect of urbanization on the vegetation health is clearly shown by the result in Figure 7 and this calls for mitigation action against the urban heat island by sustainable urban planning (Gill *et al.*, 2007)

**Land Use Changes**

A significant change in land use within the study area is shown on Figure 8 and Table 2: this result was obtained from analysis of Sentinel images between the year 2015 and 2024 (a 9-year period), for which the Sentinel data is available.

As presented on Table 2 and visually observable on Figure 8, the built-up area which covers 46.53 $km^2$ of land area, that is, about 30.10% of the total area under study, has increased 9-years after in the year 2024 to about 61.87 $km^2$ (about 40.02% of the total area under study). The increase in built-up area correspond to increase in human population of the study area, as shown on Figure 2, demonstrating that there is need for land areas required for building projects to be used for accommodations and offices as the area experiences fast urbanization. On the other hand, trees that usually provides shades and cools the land, is found to have decreased from about 21.82 $km^2$ of land area in the year 2015 to about 9.23 $km^2$ in the year 2024, this corresponds to about 14.11% of the total land area in 2015 to about 5.97% of the total land area in 2024. According to Akbari *et al.* (2001), a great contributor to UHI is a loss of tree cover.

In addition to the above, it can be observed that increase in human population has led to increase in agricultural land used to provide food for the increasing human population. The cropland (represented by Crops on Table 2) is found to have increased from about 11.44 $km^2$ (about 7.40% of the total area) in the year 2105 to about 26.74 $km^2$ (about 17.30% of the total area). This shows that urbanization and increase in human population, either by birth or by migration, can lead to increase in demand for agricultural land to produce more food, though the work of Seto *et al.* (2011) showed that there is less direct effect of expansion in agricultural land on UHI.

**Urban Heat Island (UHI)**

The dynamics of UHI in the study area shows upward trend in UHI corresponding to increase in land surface temperature (LST), decrease in vegetation cover as indicated by the NDVI, increase in human population and urbanization of the study area from the year 1984 to the year 2024. Figure 9 shows a minimal UHI index in 1984 in contrast to the year 2024 when the UHI index has increased significantly across the study area, notable is around the FULafia permanent (PS) and take-off (TS) sites, which where not in existence in 1984.

The spatial distribution of the UHI is further illustrated on Figure 10 with the profile plot of the UHI along the line of latitude 8.5 *°North* for the years 1984 and 2024. In 1984, the distinct urban area of Lafia can be seen between the longitude 8.50 and 8.54 *°East* with the elevated UHI, this elevated UHI corresponds with the high LST and low NDVI in Figures 4 and 7 respectively, for the same longitude span. By the year 2024, more areas, such as between longitudes 8.46 and 8.48 *°East* as well as between longitudes 8.56 and 8.58 *°East*, are showing elevated UHI, these are due to urbanization of the study area. This result agrees with other researchers such as Oke (1982) and Zhou *et al.* (2014), that the UHI is enhanced during the process of urbanization by reduced vegetation cover and the use of high thermal absorbing materials used in city constructions.

The class of UHI as Weak, Middle, Strong, Stronger and Strongest, for the years 1984 and 2024, shown visually on Figure 11 indicates that the Weak and Strongest classes have decreased over the years while the Middle, Strong and Stronger classes have increased. This result is quantified on Tables 3 and 4, where it can be seen that the Weak UHI progressively decrease from 79.57 *km$^2$* (about 51.27% of the study area) in 1984 to about 74.37 *km$^2$* (about 47.92% of the study area) in 2024, meanwhile the Strong UHI increased from about 16.76 *km$^2$* (about 10.80% of the study area) in 1984 to about 19.54 *km$^2$* (about 12.59% of the study area) in 2024, according to Naim & Kafy (2021) this indicates that gradual urbanization of Lafia is leading to more pronounced UHI in Lafia and its environs. As observed on Tables 3 and 4, Strong, Stronger and Strongest UHI actually peaks in 2003, this type of variation in UHI, according to Zhou *et al*. (2014), this can be linked to specific urbanization phases or climate change.

**UHI Mitigation Strategies and Recommendations**

Mitigating the UHI effect in Lafia, a sprawling tropical city and similar such cities across the globe requires multipronged and with the observed upward trend in LST and UHI, coupled with the observed lower trend in vegetation cover, in this study it is pertinent that there is a need for, among other things, use of innovative materials in urban constructions, creation of green areas across the city and adoption of effective urban planning strategy. Some of the mitigation strategies suggested against UHI are:

**Green Spaces and City Parks:** Following the observed loss of vegetation in Lafia, it is recommended that the city authority embarks on aggressive and sustained tree planting across the city streets and its environs, reserved green area and community gardens should also be created to create habitat for urban wildlife and to improve the air quality of the city (Gill *et al*., 2007). This is similar to the mitigation strategy suggested by Ebuga *et al.* (2021) for Lafia city.

**Modern Urban Planning and Community Engagement**: City planning authority should adopt a strategy that balanced urbanization with the preservation of agricultural land and natural vegetation, building codes should be made to encourage green spaces, cool roofs and tree planting with the building premises. Community education and outreach should be planned to sensitize the urban community on the effects of UHI and methods to mitigate them (Seto *et al.,* 2011).

**Urban Construction Materials**: The type of materials used for constructions in urban area should be regulated. Use of reflective roofs together with light-coloured materials and walls can reduce thermal absorbance by buildings Proper insulations and building orientation can also reduce thermal absorbance by buildings thus reducing need for air conditioning systems which in turn will reduce demand for energy consumption (Akbari *et al.,* 2001).

**Conclusion**

A detailed and comprehensive studies of the urban heat island effect in Lafia city, Nasarawa State of Nigeria, together with the impact of land surface temperature variation and vegetation cover degradations as a result of Lafia urbanization. The link between city expansion and increasing urban heat island effect, rise in surface temperature and vegetation degradation has been shown, this has significant effect on the local climate, local ecosystem health as well as human well-being. The findings of this study necessitate strategies for mitigating UHI in Lafia city and other similar sprawling cities across the globe.

**Data Availability Statement**
All the data used for this study is available on the Google Earth Engine's public data catalog at https://developers.google.com/earth-engine/datasets/catalog

**Acknowledgement**
This research was supported by a 2025 Institutional Based Research (IBR) grant from the Tertiary Education Trust Fund (TETFund) of Nigeria. Hence, the authors wish to appreciate TETFund for the funding of this research